\documentstyle[12pt]{article}
\begin{document}
\title{ON PRIMORDIAL COSMOLOGICAL DENSITY FLUCTUATIONS IN 
 THE EINSTEIN-CARTAN GRAVITY AND COBE DATA}
\author{D. PALLE \\
Zavod za teorijsku fiziku, Institut Rugjer Bo\v {s}kovi\'{c} \\
P.O.Box 1016 Zagreb, Croatia}  
\date{ }
\maketitle
 We study cosmological density fluctuations within a covariant and
 gauge-invariant fluid-flow approach for a perfect fluid in the 
 Einstein-Cartan gravity and derive the corresponding Raychaudhuri
 type of inhomogeneous coupled differential evolution equations of
 the second order.
 It appears that the quantum fluctuations of spin trigger primordial
 density inhomogeneities at the scale of weak interactions. 
 These inhomogeneities
 are then evolved precisely to the value measured by COBE mission
 at the scale of decoupling.
\\

\section{Introduction}

As well as 
cosmography and nucleosynthesis, structure formation nowadays 
represents the most important part of our theoretical 
comprehension of the Universe \cite{Kolb}.
Cosmological models with cold and hot dark matter, with or without
the cosmological constant, can fairly well describe the formation and evolution
of cosmological structures at small and large scales in the
Universe, mostly within the Friedmann-Lema$\hat{i}$tre-Robertson-Walker
metric models of the Einstein gravity \cite{Kolb}.
However, the milestone assumption for the structure formation is that certain
small density inhomogeneities cause the growth into large
inhomogeneities observed today.
These small primordial density inhomogenities were observed in 1992 by COBE 
mission as the large-angle anisotropy of CMBR \cite{COBE} .

In this paper we want to show that it is possible to solve the problem of
the primordial density inhomogeneity within the Einstein-Cartan(EC) gravity
without referring to the dynamics of the cosmological scalar field.

In the next section we derive the evolution equations for the density
contrast within the Hawking fluid-flow approach\cite{Hawk} 
and with the covariant and
gauge-invariant variables of Ellis et al.\cite{Ellis} in the EC 
gravity with perfect fluid described by Obukhov and Korotky\cite{Obuh}.

The last section is devoted to the solution of inhomogeneous evolution
equations for the spacetime with expansion, acceleration and torsion,
as well as to the estimates of the density contrast at various scales,
including the scale of decoupling of CMBR.

\section{Inhomogeneous coupled evolution equations for the density contrast}

The theory of small fluctuations in general relativity started with the
work of Lifshitz \cite{Lif} within a coordinate approach and it was
formulated using the gauge-invariant variables by Bardeen \cite{Bard}.

Our task to study the fluctuations in a more general spacetime with
nonvanishing expansion, acceleration, vorticity, shear and torsion,
requires a different, more elegant and powerful approach, such as the
fluid-flow formalism \cite{Hawk} supplied with covariant and 
gauge-invariant variables \cite{Ellis}.
 
We start with the formulation of perfect fluid in the EC
gravity described by Obukhov and Korotky\cite{Obuh} . 
Definitions, field equations and conservation equations, as a consequence
of the Bianchi identities, look as it follows\cite{Obuh}:

\begin{eqnarray}
\tilde{\Gamma}^{\alpha}_{\beta \mu}&=&\Gamma^{\alpha}_{\beta \mu}+
Q^{\alpha}_{\cdot \beta \mu}+Q_{\beta \mu \cdot}^{\ \ \alpha}+
Q_{\mu \beta \cdot}^{\ \ \alpha},
  \hspace*{32 mm} \nonumber \\
\Gamma^{\alpha}_{\beta \mu}&=&\frac{1}{2}g^{\alpha \nu}
(\partial_{\beta}g_{\mu \nu}+\partial_{\mu}g_{\beta \nu}
-\partial_{\nu}g_{\beta \mu}), \hspace*{6 mm} \nonumber 
\end{eqnarray}

\begin{eqnarray}
\tilde{R}^{\alpha}_{\cdot \beta \mu \nu}=\partial_\mu\tilde{\Gamma}
^{\alpha}_{\beta \nu}-\partial_{\nu}\tilde{\Gamma}^{\alpha}
_{\beta \mu}+\tilde{\Gamma}^{\alpha}_{\gamma \mu}\tilde{\Gamma}
^{\gamma}_{\beta \nu}-\tilde{\Gamma}^{\alpha}_{\gamma \nu}
\tilde{\Gamma}^{\gamma}_{\beta \mu}, \hspace*{19 mm} \nonumber \\
\tilde{\Gamma}_{\beta \gamma}^{\alpha}\rightarrow\Gamma_{\beta \gamma}
^{\alpha}\Rightarrow\tilde{R}^{\alpha}_{\cdot \beta\gamma\delta}
\rightarrow R^{\alpha}_{\cdot \beta\gamma\delta},
\tilde{\nabla}_{\mu}\rightarrow\nabla_{\mu},
 \hspace*{25 mm} \nonumber \\
Q^{\alpha}_{\cdot\beta\mu}=\kappa u^{\alpha}S_{\beta\mu},\ 
\kappa=\frac{8\pi G_{N}}{c^{4}},\ u^{\mu}S_{\mu\nu}=0,
\ S^{2}\equiv\frac{1}{2}S^{\mu\nu}S_{\mu\nu},\hspace*{10 mm} \nonumber \\
R_{\mu\nu}-\frac{1}{2}g_{\mu\nu}R=\kappa T^{eff}_{\mu\nu},
\hspace*{40 mm}  \\
T^{eff}_{\mu\nu}=-p_{eff}g_{\mu\nu}+u_{\mu}u_{\nu}(p_{eff}+
\rho_{eff})-2(g^{\alpha\beta}+u^{\alpha}u^{\beta})
\nabla_{\alpha}[u_{(\mu}S_{\nu )\beta}],\nonumber \\
(\alpha\beta )=\frac{1}{2}(\alpha\beta+\beta\alpha ),\ 
p_{eff}=p-\kappa S^{2}-\Lambda,\ \rho_{eff}=\rho-\kappa S^{2}+
\Lambda, \nonumber \\
u^{\mu}\nabla_{\mu}\rho=-(\rho +p)\nabla_{\mu}u^{\mu},\hspace*{45 mm} \\
(\rho+p)a_{\mu}+(-\delta^{\nu}_{\mu}+u^{\nu}u_{\mu})\nabla_{\nu}p
+2\tilde{\nabla}_{\nu}(u^{\nu}a^{\alpha}S_{\mu\alpha})+
S_{\alpha\beta}\tilde{R}^{\alpha\beta}_{\cdot\cdot\mu\nu}u^{\nu}=0, \\
a_{\mu}\equiv u^{\nu}\tilde{\nabla}_{\nu}u_{\mu}. \hspace*{50 mm} \nonumber 
\end{eqnarray}

To derive the Raychaudhuri type of evolution equations for expansion
and vorticity and the corresponding constraint equations, we contract
Ricci identities by various tensor structures that contain the four-velocity and
the projector orthogonal to the four-velocity \cite{Hawk,Hehl}:

\begin{eqnarray}
(\tilde{\nabla}_{\mu}\tilde{\nabla}_{\nu}-\tilde{\nabla}_{\nu}
\tilde{\nabla}_{\mu})u_{\lambda}=-\tilde{R}^{\sigma}_{\cdot
\lambda\mu\nu}u_{\sigma}-2Q^{\sigma}_{\cdot\nu\mu}\tilde{\nabla}_{\sigma}
u_{\lambda}, \hspace*{20 mm} 
\end{eqnarray}

\begin{eqnarray}
u^{\mu}u_{\mu}&=&1,\ h_{\mu\nu}\equiv g_{\mu\nu}-u_{\mu}u_{\nu},
\nonumber \\
\Theta&\equiv&\nabla_{\mu}u^{\mu},\ a_{\mu}\equiv\dot{u}_{\mu}, 
\nonumber \\
\sigma_{\mu\nu}&\equiv&[\frac{1}{2}(\nabla_{\beta}u_{\alpha}+
\nabla_{\alpha}u_{\beta})-\frac{1}{3}\Theta h_{\alpha\beta}]
h^{\alpha}_{\cdot\mu}h^{\beta}_{\cdot\nu},  \nonumber\\
\tilde{\omega}_{\mu\nu}&\equiv&\frac{1}{2}(\tilde{\nabla}_{\beta}u_{\alpha}
-\tilde{\nabla}_{\alpha}u_{\beta})h^{\alpha}_{\cdot\mu}h^{\beta}_{\cdot\nu},
 \nonumber\\
\tilde{\nabla}_{\mu}&\rightarrow&\nabla_{\mu}\Rightarrow
\tilde{\omega}_{\mu\nu}\rightarrow\omega_{\mu\nu}, 
\nonumber 
\end{eqnarray}

evolution equations:

\begin{eqnarray}
\dot{\Theta}&=&2\omega^{2}+2Q^{2}-2\sigma^{2}-\frac{1}{3}\Theta^{2}
+\tilde{\nabla}^{\mu}a_{\mu}-\frac{1}{2}\kappa (\rho+3p-2\Lambda),
  \\
\dot{\omega}_{\perp\mu\nu}&=&-\frac{2}{3}\Theta\omega_{\mu\nu}+
2\sigma_{\gamma[\mu}\omega_{\nu]\cdot}^{\ \gamma}
-\tilde{\nabla}_{[\lambda}\dot{u}_{\sigma ]}h^{\sigma}_{\cdot\mu}
h^{\lambda}_{\cdot\nu},  \\
\dot{\sigma}_{\perp\alpha\beta}&=&-\sigma_{\alpha\cdot}^{\ \kappa}
\sigma_{\beta\kappa}+\omega_{\alpha\cdot}^{\ \kappa}\omega_{\beta\kappa}
-\frac{2}{3}\Theta\sigma_{\alpha\beta}-\frac{1}{3}h_{\alpha\beta}
(2\omega^{2}+2Q^{2}-2\sigma^{2}+\nabla^{\mu}u_{\mu}) \nonumber \\
&+&h^{\ \mu}_{\alpha\cdot}h^{\ \nu}_{\beta\cdot}\nabla_{(\nu}\dot{u}_{\mu)}
-\dot{u}_{\alpha}\dot{u}_{\beta}, \\
\omega^{2}&\equiv&\frac{1}{2}\omega_{\mu\nu}\omega^{\mu\nu},\ 
\sigma^{2}=\frac{1}{2}\sigma_{\mu\nu}\sigma^{\mu\nu}, 
\hspace*{35 mm} \nonumber \\
\dot{\omega}_{\perp\alpha\beta}&\equiv&h_{\alpha\cdot}^{\ \gamma}
h_{\beta\cdot}^{\ \delta}u^{\epsilon}\tilde{\nabla}_{\epsilon}\omega_{\gamma
\delta},\ [\alpha \beta]=\frac{1}{2}(\alpha \beta -\beta \alpha ), \nonumber 
\end{eqnarray}

constraint\ equations:

\begin{eqnarray}
h^{\mu}_{\cdot\nu}\tilde{\nabla}_{\mu}\Theta&=&\frac{3}{2}[(\tilde{\nabla}_{\beta}
\omega^{\beta}_{\cdot\gamma}+\tilde{\nabla}_{\beta}\sigma^{\beta}_
{\cdot\gamma})h^{\gamma}_{\cdot\nu}+\dot{u}^{\beta}(\omega_{\nu\beta}+
\sigma_{\nu\beta})],  \\
\tilde{\nabla}^{\alpha}\tilde{\omega}_{\alpha}
&=&-2\tilde{\omega}_{\alpha}\dot{u}^{\alpha},\ 
\tilde{\omega}_{\alpha}=\frac{1}{2}\epsilon_{\alpha \beta \gamma \delta}
\tilde{\omega}^{\gamma \delta}u^{\beta}.
\end{eqnarray}

To derive the above equations, we also use EC field equations, where 
necessary. Notice that in the evolution equation for expansion, the term
$\omega_{\mu\nu}Q^{\mu\nu}$ is cancelled by the same term on the right-hand
side of the Ricci identity, contrary to the results
 obtained from the incomplete treatments of
previous authors\cite{Stew}.

Following Ellis et al., we introduce covariant and gauge-invariant variables
that are in fact the orthogonal spatial gradients of scalar density and
expansion, thus describing in a more natural way "a real spatial
 fluctuation, rather than a fictitious time fluctuation"\cite{Ellis}:

\begin{eqnarray}
{\cal D}_{\mu}&\equiv&R(t) \frac{^{(3)}\nabla_{\mu}\rho}{\rho}\equiv
R(t) \chi_{\mu};\ {\cal Z}_{\mu}\equiv R(t) ^{(3)}\nabla_{\mu}\Theta
\equiv R(t) Z_{\mu}, \\
R(t)&=&cosmic\ scale\ factor,\ ^{(3)}\nabla_{\mu}\equiv
h_{\mu\nu}\nabla^{\nu}. \hspace*{40 mm} \nonumber 
\end{eqnarray}

It is important to underline that these covariant variables within
the perfect-fluid model in the EC gravity are also gauge-invariant for
the metric with vorticity, acceleration and shear, because of the
time-dependence of mass density and pressure.
In the presence of vorticity and acceleration there are no more
hypersurfaces orthogonal to the fluid flow, but this is not a
deficiency because the variables are defined and interpreted locally,
 with possible further local decomposition \cite{Ellis}:

Acting on the evolution equations by the covariant derivative, 
and acknowledging the identity:

\begin{eqnarray}
^{(3)}\nabla_{\mu}\dot{f}-(^{(3)}\nabla_{\mu}f)^{\cdot}_{\perp}&=&
\dot{f}a_{\mu}+\frac{1}{3}\Theta ^{(3)}\nabla_{\mu}f
+ ^{(3)}\nabla_{\delta}f(\sigma^{\delta}_{\cdot\mu}+\omega^{\delta}
_{\cdot\mu}), \\
( ^{(3)}\nabla_{\mu}f)^{\cdot}_{\perp}&\equiv&u^{\sigma}h^{\lambda}
_{\cdot\mu}\nabla_{\sigma}(h^{\epsilon}_{\cdot\lambda}\nabla_{\epsilon}f).
 \nonumber
\end{eqnarray}

one can immediately obtain the evolution equations for the density and
expansion contrast vectors:

\begin{eqnarray}
h_{\mu\cdot}^{\ \nu}\dot{\chi}_{\nu}&=&\Theta (\frac{p}{\rho}-
\frac{1}{3})\chi_{\mu}-(\sigma^{\nu}_{\cdot\mu}+\omega^{\nu}_{\cdot\mu})
\chi_{\nu}-(1+\frac{p}{\rho})Z_{\mu}-\frac{\Theta}{\rho}J_{\mu}, \\
h_{\mu\cdot}^{\ \nu}\dot{Z}_{\nu}&=&G\dot{u}_{\mu}-B^{\nu}_{\cdot\mu}Z_{\nu}
- ^{(3)}\nabla_{\mu}C-\frac{1}{2}\kappa\rho\chi_{\mu}-\frac{3}{2}\kappa J_{\mu},
\\
G&=&-2\omega^{2}-2Q^{2}+2\sigma^{2}+\frac{1}{3}\Theta^{2}-\nabla^{\nu}
\dot{u}_{\nu}-\kappa\rho-\kappa\Lambda, \nonumber \\
B^{\nu}_{\cdot\mu}&=&\Theta\delta^{\nu}_{\mu}+\sigma^{\nu}_{\cdot\mu}+
\omega^{\nu}_{\cdot\mu}, \nonumber \\
C&=&2\sigma^{2}-2\omega^{2}-2Q^{2}-\nabla^{\nu}\dot{u}_{\nu}, \nonumber \\
J^{\mu}&=&2 S^{\mu}_{\cdot\alpha}u^{\nu}\nabla_{\nu}a^{\alpha}+
h^{\mu\sigma}S_{\alpha\beta}\tilde{R}^{\alpha\beta}_{\cdot\cdot\sigma\nu}u^{\nu}
. \nonumber
\end{eqnarray}

By direct insertion we can write coupled inhomogeneous differential
equations of the second order for the density-contrast vector:

\begin{eqnarray}
-\ddot{{\cal D}}_{\mu}&+&\alpha_{\mu\cdot}^{\ \nu}\dot{{\cal D}}_{\nu}+
\beta_{\mu\cdot}^{\ \nu}{\cal D}_{\nu}+\gamma_{\mu}=0,\hspace*{33 mm}\\
{\cal D}&\equiv&(-{\cal D}_{\mu}{\cal D}^{\mu})^{\frac{1}{2}},\ 
\dot{{\cal D}}_{\mu}\equiv u^{\nu}\nabla_{\nu}{\cal D}_{\mu},\ 
\ddot{{\cal D}}_{\mu}\equiv (\dot{{\cal D}}_{\mu})^{\cdot}, \nonumber 
\end{eqnarray}

\begin{eqnarray}
\alpha_{\mu\cdot}^{\nu}&=&[\frac{\dot{w}}{1+w}+\Theta (w-\frac{1}{3})+
2\frac{\dot{R}}{R}]\delta^{\nu}_{\mu}-\sigma^{\nu}_{\cdot\mu} 
-\omega^{\nu}_{\cdot\mu}-2u_{\mu}\dot{u}^{\nu}-B^{\nu}_{\cdot\mu},
  \\
\beta_{\mu\cdot}^{\ \nu}&=&[-\frac{\Theta\dot{w}(w-\frac{1}{3}}{1+w}
-\frac{\dot{w}}{1+w}\frac{\dot{R}}{R}+\dot{\Theta}(w-\frac{1}{3})
+\dot{w}\Theta      
-\Theta (w-\frac{1}{3})\frac{\dot{R}}{R} \nonumber \\ 
&+& \frac{\ddot{R}}{R}
+\frac{1}{2}\kappa\rho(1+w)]\delta^{\nu}_{\mu}  
+(\frac{\dot{R}}{R}+
\frac{\dot{w}}{1+w})(\sigma^{\nu}_{\cdot\mu}+\omega^{\nu}_{\cdot\mu})
+\frac{\dot{w}}{1+w}u_{\mu}\dot{u}^{\nu}
-(\sigma^{\nu}_{\cdot\mu}+\omega^{\nu}_{\cdot\mu})
^{\cdot}       \nonumber \\
&+&\frac{\dot{R}}{R}u_{\mu}\dot{u}^{\nu}-\dot{u}_{\mu}\dot{u}^{\nu}
-u_{\mu}\ddot{u}^{\nu}+(B^{\nu}_{\cdot\mu}+u_{\mu}\dot{u}^{\nu})
(\Theta(w-\frac{1}{3})+\frac{\dot{R}}{R})  \nonumber \\
&-&(B^{\lambda}_{\cdot\mu}+u_{\mu}\dot{u}^{\lambda})(\sigma^{\nu}_{\cdot\lambda}
+\omega^{\nu}_{\cdot\lambda}+u_{\lambda}\dot{u}^{\nu})
 \\
\gamma_{\mu}&=&R[(\frac{\dot{w}}{1+w}\frac{\Theta}{\rho}-\frac{\dot{\Theta}}
{\rho}+\frac{\Theta\dot{\rho}}{\rho^{2}})J_{\mu}-\frac{\Theta}{\rho}
\dot{J}_{\mu}   
-\frac{\Theta}{\rho}(B^{\nu}_{\cdot\mu}+u_{\mu}\dot{u}^{\nu})J_{\nu}
\nonumber \\
&+&(1+w)(-G\dot{u}_{\mu}+ ^{(3)}\nabla_{\mu}C+\frac{3}{2}\kappa J_{\mu})],
\end{eqnarray}

$\ \ \ \ \ \ 
w\equiv\frac{p}{\rho},\ c^{2}_{s}\equiv\frac{dp}{d\rho}=w+\rho\frac{dw}{d\rho}=w+\rho
\frac{\dot{w}}{\dot{\rho}},$ \\  $  \hspace*{16 mm} 
\dot{w}=-(1+w)(c^{2}_{s}-w)\Theta,\ \dot{\rho}=-(p+\rho)\Theta. 
$ \\

In the next section we study these equations and explore its
observable cosmological consequences.

\section{Results and discussion}

The standard description of the Universe usually contains only the Hubble 
expansion and perfect fluid.
However, it was shown that vorticity and acceleration play a very
important role in EC cosmology \cite{Palle1}.
Namely, because of the strong binding force, the baryonic spins act
coherently at the scale of weak interactions and consequently
a fraction of the baryon mass density produces a large baryon spin density
that takes effect as a bounce force avoiding and preventing
cosmological singularity in the EC gravity precisely at the scale of weak
interactions \cite{Palle1}.
On the other hand, the spins of cold and hot dark matter can coherently
contribute to spin and torsion at spacelike infinity when all
dynamical degrees of freedom are frozen ($T_{\gamma}=0$).
At spacelike infinity it is possible to make a relationship between
basic cosmological observables within the EC gravity \cite{Palle1,Birch}:

\begin{eqnarray}
|\omega_{\infty}|=\frac{\sqrt{3}}{2}\Sigma H_{\infty},\ 
\Sigma=\frac{l}{k+l}, \hspace*{25 mm} \\
\rho_{\infty}=\frac{3}{4\pi G_{N}}H^{2}_{\infty},\ 
\Lambda=-\frac{1}{2}\rho_{\infty}. \hspace*{25 mm} 
\end{eqnarray}

However,
the consistency with the EC field equations for the perfect-fluid model
requires the vanishing of vorticity and shear for the metric
with nonvanishing expansion and acceleration
m=0, r=R \cite{Obuh}:

\begin{eqnarray}
ds^{2}=dt^{2}-R(t)^{2}(dx^{2}+ka(x)^{2}dy^{2})-r(t)^{2}dz^{2}
-2R(t)b(x)dydt, \hspace*{10 mm} \nonumber \\
b(x)=\sqrt{l}a(x),\ a(x)=Ae^{mx},\ k,l,A,m=const,  
\nonumber  \\
EC\ equations \Rightarrow r=R,\ m=0,\  
\frac{\ddot{R}}{R}=\frac{\dot{R}^{2}}{R^{2}}
\Rightarrow \sigma=\omega=0. \hspace*{20 mm} \nonumber
\end{eqnarray}

This is not a serious obstacle because vorticity is small in comparison
with expansion. One should improve the matter part of the field
equations adding
 imperfect fluid terms, cosmic magnetic field, etc., if one
wishes to develop a more detailed picture of the Universe. Anyhow,
the $\Sigma$ parameter remains constrained by the Hubble expansion-
vorticity relationship\cite{Palle1}.
The Boltzmann equation for nonrelativistic fluid ensures that the present    
mass density does not differ significantly from that at spacelike
infinity:

\begin{eqnarray}
\rho_{m}=\rho_{CDM}+\rho_{\nu}+\rho_{B}+\rho_{\gamma}, \hspace*{40 mm} \\
\rho_{\nu}=n_{\nu}(m_{\nu_{e}}+m_{\nu_{\mu}}+m_{\nu_{\tau}})+
\frac{9}{2}n_{\nu}k_{B}T_{\nu},\ p_{\nu}=3n_{\nu}k_{B}T_{\nu},
\nonumber \\
k_{B}=Boltzmann\ constant,\ n_{i} << \frac{(2\pi m_{i}k_{B}T_{i})^{3/2}}
{h^{3}}, \hspace*{10 mm} \nonumber \\
\Rightarrow \frac{\rho_{\nu}(T_{\nu ,0})-\rho_{\nu}(0 K)}
{\rho_{\nu}(T_{\nu ,0})}={\cal O}(10^{-2}),\ 
similarly\ for\ \rho_{CDM}\ and\ \rho_{B}.   \nonumber
\end{eqnarray}

Let us now look at the form of the coefficients of inhomogeneous
evolution equations for the vector density contrast in the EC gravity
with expansion and acceleration:

\begin{eqnarray}
\alpha^{\ \nu}_{\mu\cdot}&=&(\frac{\dot{w}}{1+w}+3H(w-\frac{2}{3}))  
\delta^{\nu}_{\mu}-2u_{\mu}a^{\nu}, \hspace*{10 mm} \nonumber \\
\beta_{\mu\cdot}^{\ \nu}&=&(3\frac{H\dot{w}}{1+w}+(w-\frac{1}{3})\dot{\Theta}
+2(1+3w)H^{2}+\frac{1}{2}\kappa\rho (1+w))\delta^{\nu}_{\mu} \nonumber \\
&+&(\frac{\dot{w}}{1+w}+3H(w-\frac{2}{3}))u_{\mu}a^{\nu}-a_{\mu}a^{\nu}
-u_{\mu}\dot{a}^{\nu}, \nonumber \\
\gamma_{\mu}&=&-(1+w)G a_{\mu}R(t). \nonumber
\end{eqnarray}

By direct inspection we see that, for $R/R_{0} << 1$, the terms with
covariant derivatives are suppressed in comparison with the others, thus
one can write for an approximate solution:

\begin{eqnarray}
{\cal D}_{\mu}&=&\frac{4Q^{2}}{\kappa\rho}\sqrt{l}a\dot{R}R\delta^{2}_{\mu},\ 
for\ R\simeq 10^{-16}cm, \hspace*{30 mm} \\
{\cal D}_{\mu}&=&2\sqrt{l}a\dot{R}R\delta^{2}_{\mu},\ for\  
\frac{R}{R_{0}} <  10^{-4},\ w=\frac{1}{3},  \\
{\cal D}_{\mu}&=&\frac{3}{2}\sqrt{l}\dot{R}R\delta^{2}_{\mu}, 
for\ \frac{R}{R_{0}} > 10^{-4},\ w=0, \\
Q^{2}&=&\kappa^{2}S^{2},\ S^{2}\simeq(\frac{\hbar\rho_{B}}{m_{B}})^{2}.
 \nonumber 
\end{eqnarray}
 
Searching for a small correction to this solution one has to insert the
perturbed solution $\overline{{\cal D}_{2}}+\delta {\cal D}_{2}$  to the coupled equations
for ${\cal D}_{0}$ and ${\cal D}_{2}$  components (components
${\cal D}_{1,3}$, decoupled from the source term and ${\cal D}_{0,2}$, should vanish).
Evidently, the corrections are negligible:

\begin{eqnarray}
\delta{\cal D}_{2}&\simeq & \overline{\cal D}_{2}(\frac{R}{R_{0}})^{4},\ w=\frac{1}{3},
\nonumber \\
{\cal D}_{0}&\simeq &H^{3}R(\frac{R}{R_{0}})^{4},\ w=\frac{1}{3}.
 \nonumber
\end{eqnarray}

The density contrast is then:

\begin{eqnarray}
\frac{\delta \rho}{\rho}&\equiv&{\cal D}, \hspace*{50 mm}  \nonumber \\
{\cal D}&=&\frac{4Q^{2}}{\kappa\rho}\sqrt{\Sigma}H_{0}R(t), \ 
R\simeq 10^{-16}cm,\ w=\frac{1}{3}, \nonumber \\
{\cal D}&=&\frac{3}{2}\sqrt{\Sigma}H_{0}R(t),\ \frac{R}{R_{0}} > 10 ^{-4},\ 
w=0, \nonumber \\
\dot{H}&=&0,\ H=H_{0}.  \nonumber
\end{eqnarray}

Thus, quantum fluctuations of spin trigger the mass density
inhomogeneities at the scale of weak interactions, and later on
the density contrast evolves linearly in the cosmological scale
$R$ (even in the radiation-dominated epoch, contrary to the
usual solution of the homogeneous evolution equations where it
evolves quadratically),\cite{Ellis} receiving the following value at the
decoupling of CMBR($\sqrt{\Sigma}={\cal O}(10^{-1}),\ R_{0}\simeq H_{0}^{-1}$)
\cite{Kolb}:

\begin{eqnarray}
\frac{\delta T}{T}(large\ angle;CDM\ dominance)&=&
{\cal O}(10^{-1}){\cal D}(\frac{R}{R_{0}}\simeq 10^{-3})
\nonumber \\
\Rightarrow \frac{\delta T}{T}&\sim& 10^{-5} . \nonumber
\end{eqnarray}

This result is consistent with the measurements of the COBE-DMR
\cite{COBE}.

To conclude, one can say that within the EC gravity it is possible 
to solve fundamental cosmological problems: (1) the present mass density
of the Universe: $\Omega_{m,0}\simeq 2$ \cite{Palle1} (2) the cosmological
constant problem: $\Omega_{\Lambda}\simeq -1$ \cite{Palle1}
(3) the absence of cosmological singularity: $R_{min}\simeq 10^{-16} cm$
 \cite{Palle1,Palle2} (4) the source of density inhomogeneities:
quantum fluctuations of spin (this paper), (5) strength of the primordial
density contrast (this paper).

The assumption that the physical space is not contractible
(finite scale is fixed by spin-torsion effects or weak interacions) makes a
connection between the EC gravity and the SU(3) conformal unification
scheme for gauge interactions in particle physics \cite{Palle2}
with the observed phenomenological consequences: (1) the scale of neutrino 
masses measured by the LSND and the SuperKamiokande \cite{Palle2} (2) anomalous
enhancement of the strong coupling (Tevatron and HERA) \cite{Palle3} 
(3) anomalous b-quark electroweak couplings (LEP and SLD) \cite{Palle3}.


\begin{thebibliography}{300}

\bibitem{Kolb} E. W. Kolb and M. S. Turner, {\it The Early Universe}
 (Addison-Wesley, Redwood City, 1990). \\
P. J. E. Peebles, {\it Principles of Physical Cosmology}
 (Princeton University Press, New Jersey, 1993). 
\bibitem{COBE} G. F. Smoot et al., Astrophy. J. {\bf 396} (1992), L1.
\bibitem{Hawk} S. W. Hawking, Astrophy. J. {\bf 145} (1966), 544.
\bibitem{Ellis} G. F. R. Ellis and M. Bruni, Phys. Rev. {\bf D40} (1989),
   1804.  \\
G. F. R. Ellis, J. Hwang and M. Bruni, Phys. Rev. {\bf D40} (1989), 1819.\\
G. F. R. Ellis, M. Bruni and J. Hwang, Phys. Rev. {\bf D42} (1990), 1035.
\bibitem{Obuh} Yu. N. Obukhov and V. A. Korotky, Class. Quant. Grav. 
{\bf 4} (1987), 1633.
\bibitem{Lif} E. M. Lifshitz, J. Phys. (Moscow)\ {\bf 10} (1946), 116.
\bibitem{Bard} K. Sakai, Prog. Theor. Phys.\ {\bf 41} (1969) 1461. \\
J. M. Bardeen, Phys. Rev. {\bf D22} (1980), 1882.
\bibitem{Hehl} F. W. Hehl, Gen. Rel. Grav.\ {\bf 5} (1974), 491.
\bibitem{Stew} J. Stewart and P. H\'{a}ji\v {c}ek, Nature Phys. 
Scien. {\bf 244} (1973), 96. \\
J. Tafel, Phys. Lett. {\bf A45} (1973), 341.
\bibitem{Palle1} D. Palle, Nuovo Cim. {\bf B111} (1996), 671.
\bibitem{Birch} P. Birch, Nature {\bf 298} (1982), 451. \\
 L.-X. Li, Gen. Rel. Grav. {\bf 30} (1998), 497.
\bibitem{Palle2} D. Palle, Nuovo Cim. {\bf A109} (1996), 1535.
\bibitem{Palle3} D. Palle, hep-ph/9804326.

\end{thebibliography}
\end{document}